\documentclass[
    reprint, amssymb, aps, prb, superscriptaddress
]{revtex4-2}

\usepackage{amsmath}
\usepackage{graphicx}
\usepackage{dcolumn}
\usepackage{bbm}
\usepackage{bm}
\usepackage{ulem}

\usepackage{xcolor}

\newcommand{\avg}[1]{\langle #1 \rangle}

\renewcommand{\d}{{\rm d}}
\newcommand{\D}{\mathcal{D}}

\newcommand{\pd}{\partial}
\newcommand{\mc}{\mathcal}
\newcommand{\bnabla}{\bm{\nabla}}

\newcommand{\T}{\mc{T}}

\newcommand{\rev}[1]{{\color{blue} #1}}
\renewcommand{\rev}[1]{{#1}}

\begin{document}

\title{Nonlocal Electrostatic Field Theory from Microscopic Description}

\author{V. Stepanyan}
\affiliation{Yerevan State University, Yerevan, Armenia}
\author{Y. Sh. Mamasakhlisov}
\email{y.mamasakhlisov@gmail.com}
\affiliation{Yerevan State University, Yerevan, Armenia}
\affiliation{Institute of Applied Problems of Physics, Yerevan, Armenia}
\author{A. E. Allahverdyan}
\affiliation{Alikhanyan National Laboratory (Yerevan Physics Institute), Yerevan, Armenia}

\begin{abstract}
    The study of electric fields in soft materials converges either to the study of point-like particles (local) in nonlinear theories or to the use of particles with finite sizes in nonlocal linear theories. In this work, we start from the microscopic equations of motion and construct a unified mean field Fokker-Planck equation that describes the non-equilibrium electrostatics of nonlocal nonlinear systems. We obtain a generalized Poisson-Boltzmann equation for such systems, as well as their electrostatic free energy expression. In the linear approximation, we obtain an anisotropic susceptibility in an isotropic fluid, which allows for local linear response inversion in electrostatics.
\end{abstract}

\maketitle

\section{Introduction}

The electrodynamics of condensed matter describes the interaction of electric fields with materials and relates this interaction to their macroscopic physical properties. However, conventional electrodynamic theory generally does not explicitly address the microscopic mechanisms underlying these properties. It distinguishes between external and internal charges, while the microscopic behavior of the latter is typically treated within statistical physics and thermodynamics. Consequently, material characteristics such as dielectric permittivity, conductivity, and acid–base properties are often introduced as phenomenological parameters rather than derived from the underlying molecular or electronic dynamics.

More microscopic approaches \cite{dpb1,dpb2,dpb3,buyukdagli,sparse1,sparse2,sparse3,RevModPhys.82.1887} seek to overcome this limitation by developing a unified theoretical framework. Nevertheless, these approaches raise several conceptual issues that warrant further examination and are revisited in the present work. One such issue concerns the point-particle approximation, which relies on the assumption that the characteristic dimensions of the particles are much smaller than their mean separation.
In generalized formulations of the Poisson–Boltzmann equation that seek to describe the thermodynamic and electrostatic properties of matter self-consistently, this separation of scales is reflected in the calculation of the dielectric permittivity ($\varepsilon$) from the response of molecular dipoles, which are represented as point-like entities \cite{dpb1,dpb2,dpb3}.

In other approaches \cite{buyukdagli}, no explicit separation of spatial scales is introduced. However, these theories are restricted to the linear electrostatic regime and therefore neglect nonlinear effects arising from the coupling between thermodynamic behavior and electrostatic interactions.
Some approaches \cite{sparse1,sparse2,sparse3} effectively neglect electrostatic interactions among the constituent particles of the medium, thereby restricting their applicability to the dilute-density regime. This limitation may be viewed as a consequence of the absence of a unified framework for describing internal charges and the electric field. 

In the present work, no intrinsic separation of spatial scales is assumed, and all particles are treated within a common theoretical framework. This formulation allows the theory to be extended to dense media, in which the mean interparticle separation is comparable to, or smaller than, the characteristic particle dimensions.

To develop a unified theoretical framework for electrodynamics in soft matter, we consider a general mixture of charged molecular species, including ions and extended particles, and formulate its equilibrium mean-field electrostatics. In the present work, the molecules are assumed to possess rigid internal structures. Nevertheless, the proposed formalism and methods can be extended to molecules with deformable or elastic structures.
We begin with a Fokker–Planck equation for the multicomponent mixture, derived directly from the microscopic equations of motion. A common formalism is introduced for all internal molecular charges, allowing their nonequilibrium thermodynamics and electrostatic interactions to be described self-consistently through the electrostatic potential ($\varphi$). By taking the equilibrium limit, we derive a generalized Poisson–Boltzmann equation together with a nonlinear and nonlocal electrostatic free-energy functional. Within this formulation, all internal charges are treated on an equal thermodynamic footing.
Related approaches have previously been employed to derive dipolar Poisson–Boltzmann equations in which the internal charge distributions of molecules are approximated by point dipoles \cite{dpb1,dpb2,dpb3}. The generalized Poisson–Boltzmann equation developed here, together with its underlying formalism, may therefore be regarded as a natural microscopic extension of macroscopic electrostatics.
Within the linear-response approximation, the electric susceptibility and dielectric permittivity emerge directly from the thermodynamic description rather than being introduced as prescribed material parameters. These quantities depend on the internal structures of the different molecular species and, in general, on the applied external electric field. The resulting equation is then solved for a simple mixture of dipolar molecules and ions to demonstrate how molecular structure modifies Debye screening.
Despite the macroscopic isotropy of the medium, the nonlocal electrostatic response gives rise to an anisotropic susceptibility. In particular, the effective local susceptibility may undergo a sign reversal, producing regions or modes in which the polarization and electric-displacement vectors are oppositely directed. This behavior originates from the nonlocal character of the dielectric response.

\section{Formalism.}
    We want to consider a fluid solution containing different rigid particles (molecules / ions). First, let us write a formalism to describe a single rigid particle, which will later make the calculations on the entire system easier to follow. Denote each type of particle with the index $\alpha$. Denote the positions of the centroids $\bm{R}_\alpha^l$ for each particle $l$ of type $\alpha$ (to be more precise, these are the viscosity weighted centriods; see \eqref{app-eq:centroid} in Appendix \S A). The orientations of the particles are given via $\bm\Omega$ ZYZ Euler angles
    \begin{equation}\begin{gathered}
        \mc{R}(\bm\Omega) = \mc{R}_z(\Omega_1)\mc{R}_y(\Omega_2)\mc{R}_z(\Omega_3),
    \end{gathered}\end{equation}
    where $\mc{R}_y,\mc{R}_z$ are rotations around the $y, z$ axes respectively. We define the molecular charge structure operator for each type of molecules $\T_\alpha(\bm\Omega)$, which acting on $\delta(\bm R - \bm x)$ gives the charge density of a single $\alpha$ type molecule located at $\bm R$ in the orientation $\bm\Omega$.\begin{equation}\label{eq:operator-def}
        \T_\alpha(\bm\Omega)_x = \int q_\alpha(\bm r) e^{-(\mc{R}(\bm\Omega)\bm{r})\bnabla_x}\d^3r,
    \end{equation}
    where $\bnabla$ is the derivative by $\bm x$ and $q_\alpha(\bm r)$ is the charge density of a type $\alpha$ particle positioned in the origin with $\bm\Omega=0$. In general, $q_\alpha$ would depend on the electric field around the particle, making the operator $\T$ non-homogeneous, i.e., it would depend on the value of $\bm x$. Thankfully, for rigid particles this is not the case. Moving forward, we will omit the coordinate subscript $x$ where the action of $\T$ is self evident. To extend this formalism to non-rigid molecules we would need a Hamiltonian for their internal degrees of freedom. The adjoint operator of $\T$ is given by
    \begin{equation}
        \T_\alpha^\dagger(\bm\Omega) = \int q_\alpha(\bm r) e^{(\mc{R}(\bm\Omega)\bm{r})\bnabla}\d^3r, 
    \end{equation}
    and the charge density of the total system can be written
    \begin{gather}
        \label{eq:total-charge-density-def}
            \rho(\bm x| \{\bm \zeta \}) = \sum_\alpha \rho_\alpha(\bm x| \{\bm \zeta \}),\\
        \label{eq:charge-density-def}
            \rho_\alpha(\bm x| \{\bm \zeta \}) = \sum_{l=1}^{N_\alpha}\T_\alpha(\bm\Omega_\alpha^l)_x\delta(\bm{R}_\alpha^{l}-\bm x),
    \end{gather}
    where $\{\bm\zeta\}$ is a point in the state space of the system given by $\bm\zeta_\alpha^l = [\bm R_\alpha^l, \bm\Omega_\alpha^l]^T$ with $N=\sum_\alpha N_\alpha$ particles. The energy of a type $\alpha$ molecule in location $\bm{R}$ with orientation $\bm\Omega$ in an external electrostatic field potential $\psi(\bm{x})$ will be by definition of the adjoint
    \begin{equation}\label{eq:operator-with-potential}
        V_{\alpha:\,\rm ext}(\bm \zeta) \equiv \int \psi(\bm x)\T_\alpha(\bm\Omega)_x\delta(\bm R - \bm x)\d^3x  = \T_\alpha^\dagger(\bm\Omega)_{R}\psi(\bm R).
    \end{equation}
    Same way, we can find the energy of a Coulomb interaction between a pair of molecules $\alpha, \bm{R}, \bm\Omega$ and $\alpha', \bm{R}', \bm\Omega'$
    \begin{equation}\begin{gathered}
         \int \frac{\mc{V}(\bm x-\bm y)}{\varepsilon_0} \T_{\alpha'}(\bm\Omega')_y\delta(\bm R' - \bm y) \T_\alpha(\bm\Omega)_x\delta(\bm R - \bm x) \d^3x\d^3y\\
         = \frac{1}{\varepsilon_0} \T_\alpha^\dagger(\bm\Omega)_R\T_{\alpha'}^\dagger(\bm\Omega')_{R'}\mc{V}(\bm R-\bm R') \equiv V_{\alpha\alpha'}(\bm\zeta,\bm\zeta'),\label{eq:interaction_pot}
    \end{gathered}\end{equation}
    where the Coulomb potential is defined as
    \begin{equation}\label{eq:coulomb-def}
        \mc{V}(\bm x - \bm y)= \frac{1}{4\pi|\bm{x}-\bm{y}|},\quad \Delta \mc{V} = -\delta(\bm{x}-\bm{y}).
    \end{equation}

\section{The Rotational Fokker-Planck equation}
    To obtain the mean field equations of our fluid solution, we start with the dynamical description of the system as a whole. For interacting rigid particles in state $\{\bm\zeta\}$ the Langevin equation is written (see Appendix \S A for the derivation)
    \begin{equation}
    \label{lange1}
        M_\alpha(\bm \zeta_\alpha^l)\dot{\bm \zeta}_\alpha^l = -\bnabla_\alpha^l \mc{U} + \bm \eta_{\alpha l},
    \end{equation}
    where the interaction potential is
    \begin{equation}\label{eq:true-energy}
        \mc{U}(\{\bm \zeta\}) = \sum_{\alpha l} \bigg[V_{\alpha:\,\rm ext}(\bm \zeta_\alpha^l) + \frac{1}{2}\sum_{\alpha' k} V_{\alpha\alpha'}(\bm\zeta_\alpha^l,\bm\zeta_{\alpha'}^k)\bigg],
    \end{equation}
    and $\bm \eta$ is a normal distributed stochastic noise with
    \begin{equation}
        \avg{\bm \eta_{\alpha l}} = 0 \quad
        \avg{\eta_{\alpha l}^\mu(t)\eta_{\alpha l}^\nu(t')} = 2TM_\alpha^{\mu\nu}(\bm \zeta_\alpha^l(t))\delta(t - t'),
    \end{equation}
    with $T$ the temperature at $k_{\rm B}=1$. The explicit formula for matrix $M_\alpha(\bm\zeta)$ is found in \S\ref{app:Langevin}. It, however, is not essential for our results and does not affect the equilibrium state. The $\bnabla_\alpha^l$ is a 6 dimensional cartesian gradient. 

    The Langevin equation (\ref{lange1}) describes the influence of a thermal bath (via friction and noise) on a single molecule modeled as a rigid body. The interaction between different molecules is naturally modeled via Coulomb interaction. As is well known in plasma physics \cite{ll10} (see also \cite{santangelo} for a recent, related discussion), the long-range part of that interaction can be described via the mean-field. The collisional part of the Coulomb interaction leads to the Boltzmann equation, or -- for sufficiently dense gases and liquids -- to Langevin and Fokker-Planck equations; cf.~(\ref{lange1}).
    
    The corresponding Fokker-Planck equation for the joint probability density $P(\{\bm\zeta\},\, t)$ reads from (\ref{lange1}, \ref{app-eq:fokker-planck})
    \begin{equation}\label{eq:Fokker-Planck}
        \frac{\partial P }{\partial t}= \sum_{\alpha l} \bnabla_\alpha^l\Big[M_\alpha^{-1}(\bm\zeta_\alpha^l)\Big(P\bnabla_\alpha^l \mc{U} + T\bnabla_\alpha^lP\Big)\Big].
    \end{equation}
    Integrating \eqref{eq:Fokker-Planck} over all the $\bm \zeta$ except for one with type $\alpha$ we get
    \begin{equation}\begin{gathered}\label{eq:FP-intgrl}
            \frac{\partial P_\alpha(\bm \zeta) }{\partial t}= 
            \bnabla \bigg[ M^{-1}_\alpha(\bm \zeta) \bigg(P_\alpha(\bm \zeta)\bnabla V_{\alpha:\,\rm ext}(\bm\zeta) + T\bnabla P_\alpha(\bm\zeta)  +\\
            + \sum_{\alpha'}(N_{\alpha'}-\delta_\alpha^{\alpha'}) P_\alpha(\bm \zeta)\int \D\zeta'\, P_{\alpha'\alpha}(\bm \zeta'|\bm \zeta)\bnabla V_{\alpha\alpha'}(\bm \zeta, \bm\zeta') \bigg)\bigg].
    \end{gathered}\end{equation}
    The equation \eqref{eq:FP-intgrl} describes the exact dynamics of a fluid of rigid particles. Applying the mean field approximation amounts to the substitution
    \begin{equation}
        P_{\alpha'\alpha}(\bm \zeta'|\bm\zeta;t)=P_{\alpha'}(\bm\zeta;t) \quad\quad \forall \;\alpha,\,\alpha'.
    \end{equation}
    Hence assuming $N_{\alpha'} \gg 1$ (i.e. $N_{\alpha'} - \delta_\alpha^{\alpha'} \approx N_{\alpha'}$) and using $\D\zeta' = \d^3x\D\Omega$ with  $\int\D\Omega=\iiint_0^{2\pi,\pi,2\pi}\sin\Omega_2\d\Omega_1\d\Omega_2\d\Omega_3$ we can rewrite eq. \eqref{eq:FP-intgrl} as
    \begin{equation}\label{eq:MFFP}
        \frac{\partial P_\alpha }{\partial t}= 
        \bnabla\bigg[M^{-1}_\alpha\bigg( 
        P_\alpha(\bm\zeta;t)\bnabla U_\alpha(\bm\zeta,t)
        +T\bnabla P_\alpha(\bm\zeta;t)\bigg)\bigg],
    \end{equation}
    where $U_\alpha$ is the effective self consistent mean field potential acting on each particle $\alpha$ given by
    \begin{equation}\begin{gathered}
        \label{eq:MFP}
        U_\alpha(\bm x,\bm\Omega,t) = \T_\alpha^\dagger(\bm\Omega)\bigg[\psi(\bm x, t) +\sum_{\alpha'}\frac{N_{\alpha'}}{\varepsilon_0}\times\\
        \times\iint\mc{V}(\bm x - \bm y)\T_{\alpha'}(\bm\Omega')P(\bm y,\bm\Omega';t)\d^3y\D\Omega'\bigg].
    \end{gathered}\end{equation}
    Evidently, a separation of variables can be done in the RHS of \eqref{eq:MFP}, which, thanks to the approximation $N_\alpha \gg 1$, also decouples the particle type index from the coordinate. This allows us to decouple the rotational coordinates and the type of the molecule from the positional coordinates for all potentials $U_\alpha(\bm x,\bm\Omega,t) = \T_\alpha^\dagger(\bm\Omega)\varphi(\bm x, t)$. From \eqref{eq:operator-with-potential} we can see that the operator $\T_\alpha^\dagger$ acting on a potential gives the energy of the type $\alpha$ particle in that potential, following this similarity $\varphi(\bm x)$ represents the effective mean field electric potential of the system.
    Using \eqref{eq:coulomb-def} we can find this effective potential $\varphi$ by solving the following equation where $\Delta$ is the Laplace operator on $\bm x$ 
    \begin{equation}
        \Delta\varphi(\bm x,t) = \Delta\psi(\bm x, t) - \sum_\alpha\frac{N_\alpha}{\varepsilon_0Z_\alpha}\int\T_\alpha(\bm\Omega)P_\alpha(\bm x,\bm\Omega,t)\D\Omega.
    \end{equation}
    Thus we obtain the Poisson equation for the electric potential $\varphi(\bm x, t)$ where the probability density $P_\alpha(t)$ is a functional of $\varphi(\tau < t)$ found from the Fokker-Planck equation \eqref{eq:MFFP}.
    
\section{The Mean-Field Equilibrium }
    Assuming now that the external field $\psi(\bm x)$ is static and the stationary solution of \eqref{eq:MFFP} holds the detailed balance, we find the equilibrium distribution of one type $\alpha$ particle 
    \begin{equation}\begin{gathered}\label{eq:detailed-balance}
        \mc{P}_\alpha(\bm x, \bm\Omega)=\frac{1}{Z_\alpha}e^{-\beta \T_\alpha^\dagger(\bm\Omega)\varphi(\bm x)},\quad \beta=1/T,\\
        Z_\alpha = \iint e^{-\beta\T_\alpha^\dagger(\bm\Omega)\varphi(\bm{x})} \d^3x\D\Omega,
    \end{gathered}\end{equation}
    where $\varphi(\bm x)$ is the equilibrium electrostatic field and
    \begin{equation}\label{eq:general-poisson-boltzmann}
        \Delta\varphi = \Delta\psi - \sum_\alpha \frac{N_\alpha}{\varepsilon_0Z_\alpha}\int \T_{\alpha}(\bm\Omega)e^{-\beta\T_\alpha^\dagger(\bm\Omega)\varphi}\D\Omega.
    \end{equation}
    This is the generalized Poisson-Boltzmann equation (GPBE) for a solution of charged particles with rigid structure. It is a nonlinear nonlocal mean field (Vlasov's) equation \cite{vlasov, bavaud}. Here $Z_\alpha$ is the partition function of a single particle of type $\alpha$ interacting with the field $\varphi$.

    \rev{Since \eqref{eq:general-poisson-boltzmann} represents a nonlinear higher order moment theory, there are some physical properties of local theories that no longer hold. First, it is no longer trivial to show that the response of the system to an external electric field has a charge conjugation parity, i.e., changing the sign of the external electric field does not change the susceptibility of the system. There are in fact such conditions for which the opposite is true. The susceptibility changes when changing the sign of the external electric field, breaking charge parity of this theory. This effect is discussed in more detail in \cite{our-pre}. Another effect, due to the nonlocality of the theory, is an inversion of the local susceptibility of the system. This effect is described in section \ref{sec:lin-resp-inv} of this paper.}
    
    \subsection{Electrostatics}
    
        The potentials $\psi$ and $\varphi$ represent the electric fields
        \begin{gather}
            \bm E = -\bnabla\varphi,\\
            \bm E_{\rm ext} = -\bnabla\psi.
        \end{gather}
        From the Gauss law using \eqref{eq:general-poisson-boltzmann} the polarization equals 
        \begin{equation}\begin{gathered}\label{eq:polarization}
            \bnabla\bm P = - \rho = \varepsilon_0\Delta(\varphi - \psi) \implies\\
            \implies \bm P = \varepsilon_0\bm E_{\rm ext} - \varepsilon_0\bm E + \bnabla\times\bm{\mc{Q}},
        \end{gathered}\end{equation}
        where $\bm{\mc{Q}}$ is an arbitrary vector field. Using \eqref{eq:polarization} we also get the displacement
        \begin{equation}
            \bm D = \varepsilon_0\bm E_{\rm ext} + \bnabla\times\bm{\mc{Q}},
        \end{equation}
        giving rise to the equation
        \begin{equation}
            \bm D = \bm P + \varepsilon_0\bm E.
        \end{equation}
        \rev{The value of $\bm{\mc Q}$ can in theory be found from \eqref{eq:polarization}. Since $\bm E$ and $\bm E_{\rm ext}$ are both gradients one can apply the Helmholtz decomposition theorem or take the curl of the equation to get
        \begin{equation}\label{eq:helmholtz-decomp}
            \bnabla\times\bm P = \bnabla(\bnabla\bm{\mc{Q}}) - \Delta \bm{\mc Q},
        \end{equation}
        however,} since we are working with finite sized discrete objects, we do not have a local expression for the polarization of the system. As such, we cannot derive the value of the $\bm{\mc{Q}}$ within the limits of our theory. In the rest of the manuscript, we will take $\bm{\mc{Q}} = 0$. \rev{This approximation states that the transverse component of the polarization field vanishes and only the longitudinal component remains.}
        
        We can also find the average internal energy of the system from \eqref{eq:true-energy} using \eqref{eq:MFP}
        \begin{equation}
            \avg{\mc{U}} = \sum_\alpha \frac{N_\alpha}{2} \avg{\T_\alpha^\dagger (\varphi + \psi)}_\alpha,
        \end{equation}
        where $\avg{\cdot}_\alpha$ is averaging over the equilibrium probability distribution \eqref{eq:detailed-balance}. The free energy of the system can then be written
        \begin{equation}\begin{gathered}\label{eq:free-energy}
            F = \avg{\mc{U}} + T\sum_\alpha N_\alpha \avg{\ln\mc{P_\alpha}}_\alpha =\\
            =\sum_\alpha N_\alpha\bigg[\frac{1}{2}\avg{\T_\alpha^\dagger(\psi - \varphi)}_\alpha - T\ln {Z_\alpha}\bigg] = \\
            =\frac{\varepsilon_0}{2}\int (\psi-\varphi)\Delta(\psi-\varphi)\d^3x - T\sum_\alpha N_\alpha\ln{Z_\alpha},
        \end{gathered}\end{equation}
        where the last line is obtained via integration by parts and substitution of \eqref{eq:general-poisson-boltzmann}. Assuming the electric fields rapidly approach zero near the boundary/infinity, we can write the free energy \eqref{eq:free-energy} as
        \begin{equation}\label{eq:free-energy-elstat}
            F=-\frac{1}{2\varepsilon_0}\int \bm P^2\d^3x - T\sum_\alpha N_\alpha\ln{Z_\alpha}.
        \end{equation}
    Equation \eqref{eq:free-energy} is the nonlinear form of free energy as the value of Vlasov's functional at its minimizer \eqref{eq:general-poisson-boltzmann}. The expression accounts for the polarization that lowers the total energy of the system and for the thermal disordering effects that drive the particles' random orientation.

\section{Linearization}

    The GPBE \eqref{eq:general-poisson-boltzmann} is a highly nonlinear nonlocal equation which are commonly only solvable via iterative numerical calculations \cite{driver}. Nevertheless, the linear approximation of the GPBE can be solved analytically. To get linear equations for $\varphi$ we assume that it is a smooth function with small values $|\varphi|\ll 1$, $|\bnabla\varphi|\ll 1$ etc and plug it back into \eqref{eq:general-poisson-boltzmann}
    \begin{equation}\label{eq:linearized-gpb}
        \Delta\varphi = \Delta\psi + \sum_\alpha\frac{\beta c_\alpha}{8\pi^2\varepsilon_0} \int\T_{\alpha}(\bm\Omega)\T^\dagger_\alpha(\bm\Omega)\varphi\D\Omega,
    \end{equation}
    where $c_\alpha=N_\alpha/V$ is the average concentration of the type $\alpha$ molecule, $V$ is the volume of the system, the system is not charged $\bigg[\sum_\alpha c_\alpha \T_\alpha\bigg] 1 = 0$, and the $\int\D\Omega = 8\pi^2$ has been used. Applying a Fourier transform to equation \eqref{eq:linearized-gpb}, we get
    \begin{gather} \label{eq:fourier-gpb}
        -k^2\tilde{\varphi} = -k^2\tilde{\psi} + \sum_\alpha\frac{\beta c_j \tilde{\varphi}}{8\pi^2\varepsilon_0}\int \tilde{\T}_\alpha(\bar{\Omega},\bm{k})\tilde{\T}^*_\alpha(\bar{\Omega},\bm{k})\D\Omega, \\ 
        \tilde{\T}_\alpha(\bm\Omega, \bm{k}) = \int q_\alpha(\bm r) e^{-i\bm k\mc{R}(\bm\Omega)\bm r} \d^3r.
    \end{gather}
    The integration over the Euler angles in \eqref{eq:fourier-gpb} results in
    \begin{gather}
        \label{eq:varphi-fourier-result}
            \tilde{\varphi}(\bm k) = \frac{k^2\tilde{\psi}(\bm k)}{k^2 + \sum_\alpha\frac{\beta c_\alpha}{\varepsilon_0}s_\alpha(k)},\\
        \label{eq:structure-result}
            s_\alpha(k) = \int q_\alpha(\bm r)q_\alpha(\bm r')\,{\rm sinc}{\big[k|\bm{r}-\bm{r}'|\big]}\d^3r\d^3r',
    \end{gather}
    where the sinc function is ${\rm sinc}\,x=\sin{x}/x$. Here we have introduced the electric structure function $s_\alpha(k)$ of a type $\alpha$ molecule, which corresponds to the shift of the Fourier image of the total field potential compared to the external field potential. It is worth to note that this structure function was already obtained in \cite{buyukdagli}, where the authors derive the linear response theory for nonlocal particles from the microscopic description. The advantage of our formalism here is that it allows us to treat the ions in the solution and the rest of molecules together in a single expression.

    This structure function was discussed in \cite{sparse1,sparse2, sparse3} albeit erroneously, as the authors in their works use a different form of \eqref{eq:fourier-gpb} where the second term of RHS has $\tilde{\psi}$ instead of $\tilde{\varphi}$, which can only describe a very sparse medium where the particles do not interact with each other.
    
    Importantly $s_\alpha(k)\geq0$ as it is an integral of a nonnegative function $\tilde{\T}\tilde{\T}^*$. The structure function \eqref{eq:structure-result} can also be written using the Fourier forms $\tilde{q}_\alpha(\bm \omega)$ of charge densities $q_\alpha(\bm r)$ 
    \begin{equation}\label{eq:structure-fourier}
        s_\alpha(k)=2\pi^2\int \tilde{q}_\alpha(-\bm\omega) \tilde{q}_\alpha(\bm\omega) \frac{\delta(\omega-k)}{\omega k}\d^3\omega.
    \end{equation}

    \subsection{Linear Electrostatics}

        Taking the inverse fourier of \eqref{eq:varphi-fourier-result} noting that \eqref{eq:structure-result} is an even function we can write
        \begin{equation}\label{eq:perm-op}
            \varphi(\bm x) = \hat{\varepsilon}^{-1}\psi(\bm x), \quad\quad \tilde\varepsilon(\bm k, 0) = 1+ \sum_\alpha\frac{\beta c_\alpha}{\varepsilon_0k^2}s_\alpha(k),
        \end{equation}
        with the linear operator $\hat{\varepsilon} = \tilde\varepsilon(i\bnabla, 0)$ taking the role of dielectric permittivity. Here, $\tilde\varepsilon(\bm k, \omega)$ is the dynamic dielectric permittivity in the Fourier form, where we have found the static permittivity($\omega=0$). Importantly, the expression for permittivity \eqref{eq:perm-op} only includes even powers of $\bm k$ and therefore $\hat{\varepsilon}$ is a power series of the Laplace operator $\Delta$. We can also write \eqref{eq:perm-op} in a different, more conventional form by multiplying both sides by $\hat\varepsilon$ and then taking their gradients
        \begin{equation}\label{eq:lin-elstat-def}
            \bm D = \varepsilon_0\hat\varepsilon\bm E,\quad \bm P = \varepsilon_0\hat\chi\bm E,\quad \bm P = \hat{\chi}_{D} \bm D,
        \end{equation}
        where $\hat{\chi} = \hat{\varepsilon} - 1$ and $\hat{\chi}_{D} = \hat{\chi}\hat{\varepsilon}^{-1}$ are the electric susceptibilities of the system. We can also expand the second term of \eqref{eq:free-energy-elstat} via $\varphi$ to derive the free energy of the linearized electrostatics
        \begin{equation}\begin{gathered}
            \sum_\alpha N_\alpha\ln{Z_\alpha} \approx - \sum_\alpha \frac{\beta c}{8\pi^2}\int\T_\alpha^\dagger(\bm\Omega)\varphi(\bm{x})\d^3x\D\Omega +\\
            +N\ln{8\pi^2V}+ \sum_\alpha\frac{c_\alpha\beta^2}{16\pi^2}\int \big[\T_\alpha^\dagger(\bm\Omega)\varphi(\bm{x})\big]^2 \d^3x\D\Omega =\\
            =N\ln{8\pi^2V} + \frac{\beta}{2}\int \bm E\bm P \d^3x,
        \end{gathered}\end{equation}
        where the linear term is zero as the system is not charged, i.e., $(\sum_\alpha N_\alpha\T_\alpha) 1 = 0$. Thus, the free energy becomes
        \begin{equation}\label{eq:free-energy-elstat-lin}
            F = -\frac{1}{2\varepsilon_0}\int\bm D\bm P\d^3x - NT\ln8\pi^2V.
        \end{equation}

    \subsection{Dimer Ion Solution}
        For a simple example, let us take a solution of three different molecules $\T,\T_{+},\T_{-}$ with $c_{-}=c_{+}\equiv\frac{1}{2}c_{s}$ and
        \begin{equation}\begin{gathered}
            \T_{+}=-\T_{-}=Q,\\
            \T = qe^{-\bm{a}\bnabla}-qe^{\bm{a}\bnabla}.
        \end{gathered}\end{equation}
        Here $\T$ represents dimers (dipoles with finite size) while $\T_+$ and $\T_-$ are point ions. To apply a point dipole approximation we must take $|\bm{a}|\ll1$ and $2q|\bm{a}|=p_0$. \rev{With this, the dipole structure operator becomes
        \begin{equation}
            \T = -\bm p\bnabla,
        \end{equation}
        with the DPBE from \eqref{eq:general-poisson-boltzmann}
        \begin{equation}
            \Delta\varphi = \Delta\psi - \frac{Qc_s}{\varepsilon_0} \bigg[e^{-\beta Q\varphi} - e^{\beta Q\varphi}\bigg] + \frac{N}{\varepsilon_0Z}\int\bm p\bnabla e^{-\beta\bm p\bnabla\varphi}\D\Omega, 
        \end{equation}
        where we have used $Z_{\pm} = 8\pi^2V$, $N_{\pm} = Vc_s$. Following \cite{dpb1,dpb2,dpb3} one can integrate the third term on the RHS and in the linear approximation achieve the well known Poisson-Bolztmann equation.}
        
\begin{figure}[th]
            \centering
            \includegraphics[width=0.9\linewidth]{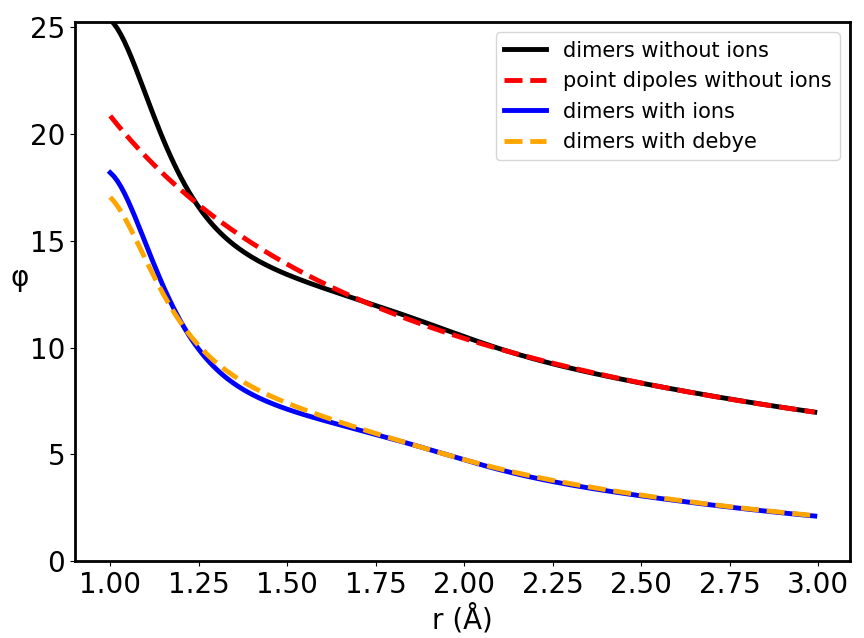}
            \caption{The profile of electric field potential in a dimer solution without ions (black, solid), point dipole solution without ions (red, dashed), dimer solution with ions (blue, solid), and dimer solution with Debye screening applied at post (yellow, dashed).}
            \label{fig:debye}
        \end{figure}

        In the linear approximation, the structure functions are then found
        \begin{equation}\begin{gathered}
                s_{+}(k)=s_{-}(k)=Q^2,\\
                s(k) = 2q^2(1-{\rm sinc}{(2ka)}) \approx p_0^2k^2/3.
        \end{gathered}\end{equation}
        Taking $\psi = \frac{\bar{Q}}{4\pi\varepsilon_0 r}$ using \eqref{eq:varphi-fourier-result} we get in the point dipole approximation
        \begin{equation}\label{eq:dipole-approx}
            \varphi(r) = \frac{\bar{Q}e^{-r/\lambda}}{4\pi\varepsilon_0\varepsilon r},\quad\varepsilon = 1 + \frac{\beta p_0^2c}{3\varepsilon_0},\quad\lambda = \sqrt{\frac{Q^2\beta c_{s}}{\varepsilon_0\varepsilon}},
        \end{equation}
        which shows the Debye screening of salt ions \cite{borukhov_1997,netz_1999,andelman-review}.
        Comparing these to the results without the point dipole approximation (using numerical inverse Fourier) in Fig.~\ref{fig:debye} we can see that the nonlocality of the particles introduce ripples in the profile of the electric potential. In Fig.~\ref{fig:debye} one can also see the electric field in a liquid without any ions to visualize the effect of screening. The values were calculated at room temperature with realistic parameter values $300\,{\rm K}$, $a = 0.5$ \AA, $Q=q=1\,{\rm e}$, $c_s = 1\,{\rm mol/l}$, and $c = 18.\bar{3}\,{\rm mol/l}$.
        
\section{Linear response inversion}\label{sec:lin-resp-inv}
    
    An interesting result of the nonlocality of \eqref{eq:lin-elstat-def} is the linear response inversion. This means that the local value of polarization is angled opposite to the displacement $\bm P\bm D < 0$. When the angle between these two is $\pi$, the local value of $\hat{\chi}_D$ is negative, which is the same as the local value $\hat{\varepsilon} \in (0, 1)$. In a more general sense, this angle can be anything between $0$ and $2\pi$, and the local value of $\hat{\chi}_D$ is a tensor.
    
    This inversion happens due to the nonlocality of the system, where the values of $\bm D$ in other parts of the system induce a polarization which affects the polarization at point $x$ and changes its direction. \rev{Effectively if the external field is weak in part {\it A} of the system but strong in part {\it B}, then the polarization induced in part {\it B} can create an electric field strong enough in part {\it A} to overpower the external field at {\it A} such that this susceptibility inversion is more free energy favorable. Therefore, the induced polarization at {\it B} amplifies the field at {\it A}. Our theory does not consider the effects of correlation and non-rigid structure of the particles, however, the intuitive reasoning of the results suggests that this "antiscreening" is caused by the internal interactions overpowering the external ones. Thus, it may be projected that correlation and nonrigidity effects on amplification should be quantitative rather than qualitative.}
    
\begin{figure}[th]
        \centering
        \includegraphics[width=\linewidth]{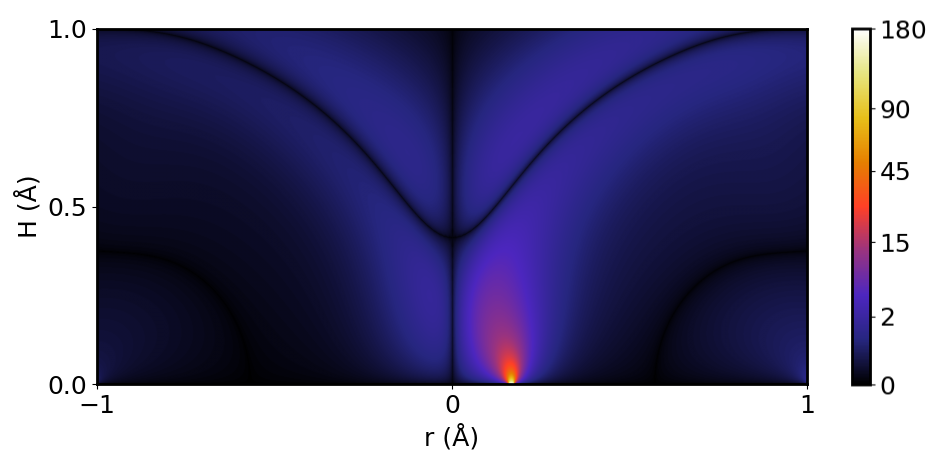}
        \includegraphics[width=\linewidth]{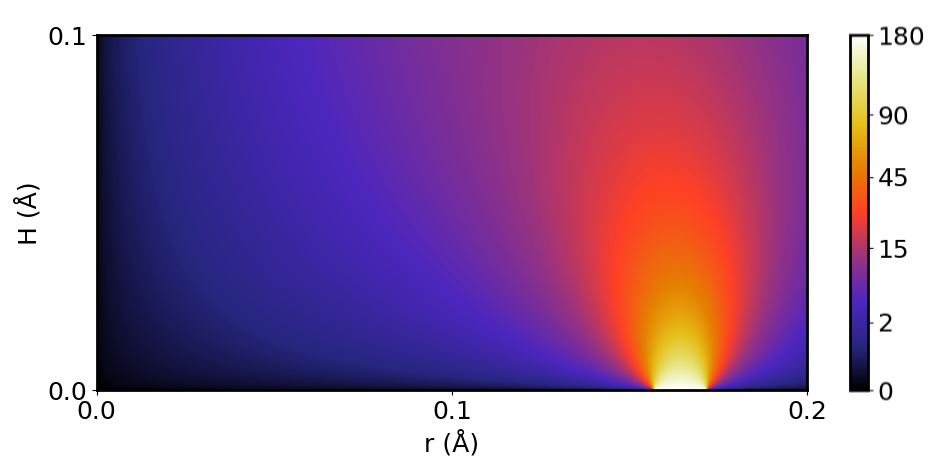}
        \caption{Angle between Displacement ($\bm D$) and Polarization ($\bm P$) for two point charges $2\,\rm e$ and $1\,\rm e$ at $2$ \AA$\,$ distance. Here $H$ is the height from the axis connecting the two charges. The rest of the parameters are $T=300\,\rm K$, $a=0.5$ \AA, $q=1\,{\rm e}$, $c = 55\,{\rm mol/l}$. The colormap is a power law with $\gamma=0.368$.}
        \label{fig:angled}
    \end{figure}

    A simple scenario where this effect can be seen is a dimer fluid with two external point charges where one of the charges is larger in value; see Fig.~\ref{fig:angled}. Here a $2\,\rm e$ charge and a $1\,\rm e$ charge are pinned at $2\,$\AA$\,$ distance between each other, $r$ is the distance from the center of the two charges in the direction of the axis, and $H$ is the height from this axis. We can see that on the axis there are points where $\bm D$ and $\bm P$ are opposite to each other (angle is $180$ degrees). In the case where these charges were both $1\,\rm e$, the maximum angle in the same segment reached $0.2$ radians, whereas if the charges were $\pm 1\,\rm e$, the maximum angle reached $0.01$ radians; both far less than even $\pi/4$ ($45$ degrees).

    This shows that even in the linear regime, an isotropic medium placed in an anisotropic external field exerts an anisotropic susceptibility. \rev{This antiscreening effect, obtained in the mean-field approximation, is partially confirmed by \textit{ab initio} thermopotentiostat molecular dynamics of nanoconfined water \cite{wipperman_2023}, in which the interface water between two electrodes is simulated, and the inverse of the transverse dielectric constant assumes values of $\varepsilon_{\perp}^{-1}>1$ directly at the interfaces. The length scale of the effect obtained above is about the same as in \cite{wipperman_2023}. The main difference between these two systems is the geometry: in our case, the effect occurs between two closely spaced point-like charges, whereas in \cite {wipperman_2023} it occurs near the interface of confined water. Our subsequent research focuses on MD simulations of the point-like ion case.}

\section{Validity of the approach}

    \rev{The mean field approximation disregards the effects of correlation on the system. A more precise equation can be achieved by adding a correlation term to \eqref{eq:MFP} like the one-loop expansion in \cite{dpb2, dpb4}. The validity of the mean field equation holds when the effects of two particle interactions are smaller than the many-body effects (like temperature $1/\beta$). Therefore, if the effective two-particle interactions, which have energy of the same order as the temperature, are rare and negligible, the mean field approximation should hold. Mathematically, this means we need to find $r = |\bm x-\bm y|$ such that the energy \eqref{eq:interaction_pot} satisfies
    \begin{equation}\label{eq:validity}
        \T_\alpha^\dagger(\bm\Omega)_x\T_\gamma^\dagger(\bm \Omega')_y \Bigg[\frac{\beta\mc{V}(\bm x -\bm y)}{\varepsilon_0\varepsilon(\bm x-\bm y)}\Bigg] \simeq 1,
    \end{equation}
    where $\varepsilon(\bm r)$ is the local value of the effective susceptibility of the system. The condition for the rarity of these collisions is then $r < c^{-1/3}$, where $c$ is the concentration of the particles. In the dimer without ion example $c^{-1/3}\approx4.8$ \AA, and in a crude approximation we can take $\varepsilon\approx27.56$ and get $r\approx3.5$ \AA, leading to $3.5 < 4.8$. While this is not a difference in orders, it still indicates a relative applicability of the method.}

    \rev{By taking rigid structures for the particles, we are ignoring the effect of induced dipolar (and higher order) moments on the particles by the electric fields. This effect has been considered for a dipolar liquid \cite{dpb4} where the assumption is that the dipoles are rod-like molecules with linear springs for bonds. We are currently working on a way to include this effect in the theory in a self consistent manner. Without this, the theory is not applicable to situations where the fields are strong enough to noticeably affect the structure of particles. To determine this limit, we can use the molecular polarizability $\kappa$ of the particles to find the effective distance $r$ at which the induced polarization is of the same order as the permanent dipole moment $\kappa \simeq ar^2\varepsilon$, where $a$ is the size of the molecule. Using $\kappa \sim a^3$ we get $r\sim a\sqrt{\varepsilon^{-1}}$. Once again, we compare $r < c^{-1/3}$ for the concentration. The value of $r$ for our dimer liquid is $r\approx 0.2$\AA, leading to $0.2 < 4.8$.}

\section{Conclusion}
    Biological systems commonly rely on electrostatic interactions between water molecules, polyelectrolytes and ions to fulfill their structural and functional properties \cite{finkel_2016,Renetal2012,Ball2017,polyelec-solutions}. Aqueous electrolyte solutions form a large portion of the systems in material science, biophysics, nanotechnology, and power cell technologies \cite{surface-forces, polyelec-solutions,polyelec-bio,dynamics,binder-battery,ion-transport-battery}. Such solutions are usually studied within the scope of the Poisson-Boltzmann (PB) mean-field equation. This approach models the behavior of point charges in a solvent with a fixed dielectric constant \cite{surface-forces}. This regime, however, is not representative of the reality of these systems. Liquids in the biological setting are commonly filled with many different types of molecules, ions, and other charged compounds, where the properties of these, such as their sizes, inhomogeneous distribution of charges, and elasticity, all play a vital role in the equilibrium electrostatics of the system.
    
    We have solved this issue by taking particles as rigid structures with a preset charge density and considering the electrostatic interactions between different molecules in the solution, while ignoring the interactions within a molecule. This approach is valid for small molecules with relatively constant structures, however, just like the standard PB equations, it can be modified to account for a large polyelectrolyte with elastic properties \cite{rudi1991, var-rudi,kumari_2024,Wu_2025}. Applying the mean field approximation to the microscopic description, we arrive to self-consistent Poisson and Fokker-Planck equations, which govern the non-equilibrium dynamics of this system. In the equilibrium, we obtain the generalized Poisson-Boltzmann equation. The resulting equation provides a unified theoretical approach to describe the electrostatic properties of a fluid without any assumptions on the dielectric permittivity of the solvent.

    \rev{Experimental measurements and computer simulations of the dielectric features of liquids remain a focus of research \cite{becker_2024}, but, to the best of our knowledge, experimental data confirming or refuting our results are currently unavailable. To address this gap, X-ray and neutron scattering studies of ion hydration structures in solution \cite{ansell_2006} or polarimetric angle-resolved second harmonic scattering on nanoparticles could be proposed \cite{bischoff_2021}. Our future research aims to add non-rigidity to this formalism, so that the predicted results can be applicable to a wider range of experimental setups.}

\section*{Acknowledgements}
    We thank Vahagn Abgaryan and Vardan Bardakhchyan for thoughtful discussions. This work was supported by the HESC of Armenia, Grants No. 21AG-1C038 and 24IRF/2-1C001.

\section*{Author Declarations}

    \subsection*{Conflict of interest}
    
    The authors have no conflicts to disclose

    \subsection*{Author Contributions}

    \textbf{V. Stepanyan:} Conceptualization (equal); Formal analysis (equal); Investigation (equal); Methodology (equal); Software (equal);  Visualization (equal); Writing - original draft (equal).

    \textbf{Y. Sh. Mamasakhlisov:} Conceptualization (equal); Formal analysis (equal); Validation (equal); Writing – review \& editing (equal).

    \textbf{A. E. Allahverdyan:} Conceptualization (equal); Methodology (equal); Software (equal); Validation (equal); Writing – review \& editing (equal).

    \subsection*{Data Availability}

    No data were created or analyzed in this study.

\appendix

\section{Rigid Body Langevin Equation}\label{app:Langevin}
    To derive the Langevin equation for a rigid body we first divide this rigid body into parts $\bm x_i$ and for each write the overdamped Langevin equation
    \begin{equation}\label{app-eq:langevin-full}
        \gamma_i \dot{\bm x_i} = -\sum_j\bnabla_i U_{ij}(|\bm x_i-\bm x_j|, t) -\bnabla_i U_i(\bm x_i, t) + \sqrt{2T\gamma_i}\bm\xi_i,
    \end{equation}
    where $\gamma_i$ is the viscosity coefficient of part $i$, $\bnabla_i$ is the gradient by $\bm x_i$, $U_{ij}$ is the potential of interaction between the parts $i,j$, $U_i$ the external potential of part $i$ and $\bm\xi_i$ is a white noise term
    \begin{equation}
        \avg{\bm \xi_i} = 0 \quad
        \avg{\xi_{i}^\mu(t)\xi_{j}^\nu(t')} = \delta_i^j\delta_\mu^\nu\delta(t - t').
    \end{equation}
    Taking the viscous weighted centroid
    \begin{equation}\label{app-eq:centroid}
        \bm x_c = \frac{1}{\gamma}\sum_i\gamma_i\bm x_i,
    \end{equation}
    where $\gamma = \sum_i\gamma_i$ and using $\bm x_i = \bm x_c + \bm r_i$ we can write
    \begin{equation}\label{app-eq:centroid-langevin}
        \gamma \dot{\bm x}_c = - \bnabla_c \sum_iU_i(\bm x_c + \bm r_i, t) + \sum_i \sqrt{2T\gamma_j} \bm\xi_i,
    \end{equation}
    where we have used $\sum_{ij}\bnabla_iU_{ij}(|\bm x_i-\bm x_j|) = 0$. With the rotation matrix $\mc{R}(\bm\Omega)$
    \begin{equation}\begin{gathered}
        \mc{R}(\bm\Omega) = \begin{bmatrix}
            \cos{\Omega_1} & -\sin{\Omega_1} & 0\\
            \sin{\Omega_1} & \cos{\Omega_1} & 0\\
            0 & 0 & 1
        \end{bmatrix}\begin{bmatrix}
            \cos{\Omega_2} & 0 & \sin{\Omega_2}\\
            0 & 1 & 0\\
            -\sin{\Omega_2} & 0 & \cos{\Omega_2}
        \end{bmatrix}\\\begin{bmatrix}
            \cos{\Omega_3} & -\sin{\Omega_3} & 0\\
            \sin{\Omega_3} & \cos{\Omega_3} & 0\\
            0 & 0 & 1
        \end{bmatrix},
    \end{gathered}\end{equation}
    the rigid rotation formalized as $\bm r_i = \mc{R}(\bm\Omega) \bm a_i$ leads to
    \begin{gather}
        \dot{\bm r}_i = L\dot{\bm\Omega} \times \bm r_i,\\\nonumber\\
        L = \begin{bmatrix}
            0 & -\sin\Omega_1 & \cos\Omega_1\sin\Omega_2\\
            0 & \cos\Omega_1 & \sin\Omega_1\sin\Omega_2\\
            1 & 0 & \cos\Omega_2
        \end{bmatrix},
    \end{gather}
    and the Langevin equation for the rotational motion can be found by cross-multiplying eq. \eqref{app-eq:langevin-full} by $\bm r_i$ and summing over $i$
    \begin{equation}\label{app-eq:rot-mid-deriv}
        IL\dot{\bm\Omega} = -\sum_i \bm r_i \times \bnabla_c U_i(\bm x_c + \bm r_{i}, t) + \sum_i \sqrt{2T\gamma_i}\bm r_i \times\bm\xi_i,
    \end{equation}
    where $I_{\mu\nu} = \sum_i \gamma_i\big[|\bm r_i|^2\delta_\mu^\nu - r_i^\mu r_i^\nu\big]$ is the "tensor of viscous inertia" and again we have no contribution of internal interactions $\sum_{ij}\bm r_i\times\bnabla_iU_{ij}(|\bm x_i-\bm x_j|) = 0$. Importantly, the first term of the LHS of \eqref{app-eq:rot-mid-deriv} can be written as $(L^T)^{-1}\bnabla_\Omega \sum_iU_i$ where $\bnabla_\Omega = [\pd_{\Omega_1},\pd_{\Omega_2},\pd_{\Omega_3}]^T$. Using this and \eqref{app-eq:centroid-langevin} with $U(\bm x_c,\bm\Omega,t) = \sum_i U_i(\bm x_c+R(\bm\Omega)\bm a_i,t)$ we can get the full Langevin equation for the six dimensional cartesian vector $\bm\zeta = [\bm x_c,\bm \Omega]^T$
    \begin{equation}\label{app-eq:langevin}
        M\dot{\bm \zeta} = - \bnabla_{\zeta} U + \bm\eta,
    \end{equation}
    where
    \begin{equation}\label{app-eq:matrix}
        M = \begin{bmatrix}
            \gamma & 0\\\\
            0 & L^TIL
        \end{bmatrix}, \quad \bm\eta = \sum_i\sqrt{2T\gamma_i}L^T\begin{bmatrix}
            \bm \xi_i\\\\
            \bm r_i\times\bm\xi_i
        \end{bmatrix}.
    \end{equation}
    The normally distributed stochastic vector $\bm \eta$ will then have the following mean and covariance.
    \begin{equation}
        \avg{\bm \eta} = 0 \quad
        \avg{\eta^\mu(t)\eta^\nu(t')} = 2TM^{\mu\nu}\delta(t - t').
    \end{equation}

\section{Rigid Body Fokker-Planck}\label{app:Fokker-Planck}

    To derive now the Fokker Planck equation we begin from the definition
    \begin{equation}
        P(\bm \zeta, t) = \avg{\delta(\bm \zeta - \bm Z[\eta;t]}.
    \end{equation}
    where $Z[\eta;t]$ is the solution of \eqref{app-eq:langevin} for a sample process $\eta$. From here we get
    \begin{equation}\begin{gathered}\label{app-eq:dtp}
        \pd_t P(\bm \zeta, t) = -\bnabla_\zeta\avg{\dot{\bm Z}\delta(\bm\zeta - \bm Z[\eta;t]} =\\
        = \bnabla_\zeta(PM^{-1} \bnabla_\zeta U) - \bnabla_\zeta [M^{-1}\avg{\bm\eta\delta(\zeta - Z[\bm\eta;t])}]
    \end{gathered}\end{equation}
    
    Let us now calculate the diffusion term. Using $\bm\eta \sim {\rm exp}{\big[-\frac{1}{4T}\int\bm\eta(t) M^{-1}\bm\eta(t)  \d t\big]}$ we get
    \begin{equation}
        \avg{F[\bm\eta;t]\eta_\mu(t)} = 2T\bigg\langle M^{\mu\nu}\frac{\delta F[\bm\eta;t]}{\delta\eta_\nu(t)}\bigg\rangle.
    \end{equation}
    And then applying it to the diffusion term
    \begin{equation}\begin{split}
        \avg{\delta(\bm\zeta-\bm Z[\bm\eta;t])\eta_\mu} =&\\= -2T(\bnabla_\zeta)_\sigma&\bigg\langle M^{\mu\nu}\frac{\delta Z_\sigma[\bm\eta;t]}{\delta \eta_\nu}\delta(\bm\zeta-\bm Z[\bm\eta;t])\bigg\rangle,
    \end{split}\end{equation}
    where we can find
    \begin{equation}
        \frac{\delta Z_\sigma[\bm\eta;t]}{\delta \eta_\nu} = \frac{1}{2}(M^{-1})^{\nu\sigma}.
    \end{equation}
    Therefore we can write for the random term
    \begin{equation}
        \avg{\delta(\bm\zeta-\bm Z[\bm\eta;t])\bm \eta} = - T\bnabla_\zeta P(\bm \zeta, t),
    \end{equation}
    and the final Fokker-Planck equation turns into
    \begin{equation}\label{app-eq:fokker-planck}
        \pd_t P(\bm \zeta, t) = \bnabla_\zeta [P(\bm \zeta, t)M^{-1}\bnabla_\zeta U + TM^{-1}\bnabla_\zeta P(\bm \zeta, t)].
    \end{equation}
    
\bibliography{dim}

\end{document}